\pdfoutput=1
\documentclass[10pt, nofootinbib, showpacs,letterpaper,aps,pra,twocolumn,superscriptaddress,floatfix]{revtex4-2}
\usepackage{blindtext}
\usepackage[english]{babel}
\usepackage[linktocpage=true, colorlinks=true, linkcolor=red, urlcolor=blue, citecolor=blue]{hyperref}
\usepackage[utf8]{inputenc}
\usepackage{pslatex}
\usepackage{setspace}
\usepackage{bbold}
\usepackage{amsthm}
\usepackage{amsmath}
\usepackage{amssymb}
\usepackage{amsfonts}
\usepackage{mathtools}
\usepackage{paralist, tabularx}
\usepackage{placeins}
\usepackage{bbm}
\usepackage{soul}
\usepackage{bm}
\DeclareMathOperator{\Tr}{Tr}

\usepackage{indentfirst}
\usepackage{mathrsfs}
\usepackage{braket}
\usepackage{graphicx}
\usepackage[margin=2cm]{geometry}
\usepackage[colorinlistoftodos]{todonotes}
\usepackage{comment}
\theoremstyle{definition}
\newtheorem{definition}{Definition}[section]

\newtheorem{proposition}{Proposition}

\newtheorem{program}{Program}
\usepackage{cancel}
\usepackage{fancyhdr}
\allowdisplaybreaks
\begin{document}
	\title{Robustness of contextuality under different types of noise as quantifiers for parity-oblivious multiplexing tasks}
	\author{Amanda M.~Fonseca}
	\affiliation{Institute of Physics, University of São Paulo, 05508-090 São Paulo, Brazil}
	\author{Vinicius P.~Rossi}
	\email{vinicius.pretti-rossi@ug.edu.pl}
	\affiliation{International Centre for Theory of Quantum Technologies, University of Gda{\'n}sk, 80-309 Gda\'nsk, Poland}
	\author{Roberto D.~Baldijão}
	\affiliation{International Centre for Theory of Quantum Technologies, University of Gda{\'n}sk, 80-309 Gda\'nsk, Poland}
	\author{John H.~Selby}
	\affiliation{International Centre for Theory of Quantum Technologies, University of Gda{\'n}sk, 80-309 Gda\'nsk, Poland}
	\affiliation{Theoretical Sciences Visiting Program, Okinawa Institute of Science and Technology Graduate University, Onna, 904-0495, Japan}
	\author{Ana Belén Sainz}
	\affiliation{International Centre for Theory of Quantum Technologies, University of Gda{\'n}sk, 80-309 Gda\'nsk, Poland}
	\affiliation{Theoretical Sciences Visiting Program, Okinawa Institute of Science and Technology Graduate University, Onna, 904-0495, Japan}
	\affiliation{Basic Research Community for Physics e.V., Germany}
	
	\begin{abstract}
		Generalized contextuality is the notion of nonclassicality powering up a myriad of quantum tasks, among which is the celebrated case of a two-party information processing task where classical information is compressed in a quantum channel, the parity-oblivious multiplexing (POM) task. The success rate is the standard quantifier of resourcefulness for this task, while robustness-based quantifiers are as operationally motivated and have known general properties. In this work, we leverage analytical and numerical tools to estimate robustness of contextuality in POM scenarios under different types of noise. We conclude that for the 3-to-1 case robustness of contextuality to depolarisation, as well as a minimisation of robustness of contextuality to dephasing over all bases, are good quantifiers for the nonclassical advantage of this scenario. Moreover, we obtain a general relation between robustness of contextuality to depolarisation and the success rate in any $n$-to-1 POM scenario and show how it can be used to bound the number of bits encoded in this task.
	\end{abstract}
	
	\maketitle
	
	Since the advent of quantum technologies, the superiority of quantum over classical methods has been anticipated across various areas, such as computation, information processing, and metrology. A celebrated result by Holevo~\cite{holevo73} suggests that communication tasks may be an exception, stating that one can never retrieve more than $n$ bits of classical information out of $n$ qubits, apparently forbidding quantum superiority over classical strategies for communicating. Quantum advantage, however, can arise in communication tasks that profit from the quantum systems' capacity to \emph{carry} more information than classical ones~\cite{gisin07,brassard03}. Therefore, such tasks prove interesting from both practical and foundational perspectives: they point to advantageous technological opportunities using quantum systems and also showcase how Holevo's theorem can be conciliated with such an advantage in communication. We here focus on one of such tasks, called \emph{parity-oblivious multiplexing} (POM), in which Alice sends Bob an $n$-bit string by sharing a smaller number of qubits. It has been demonstrated that the maximum success probability using quantum strategies for POM tasks is strictly higher than the maximum success probability obtained using classical systems, constituting an instance of quantum advantage over classical strategies for communication~\cite{spekkens08}.
	
	It was previously shown that contextuality~\cite{spekkens05} powers quantum advantage in POM tasks~\cite{spekkens08}. Besides POM, contextuality powers up other aspects of communication~\cite{saha2019,sumit23, roch22}, as well as computation~\cite{schmid22}, machine learning~\cite{bowles23}, information processing~\cite{spekkens09, ambainis19, chailloux16, yadavalli22}, metrology~\cite{lostaglio20}, state-dependent cloning~\cite{lostaglio22}, and state discrimination~\cite{schmid18,shin21,flatt22,mukherjee22}. Contextuality first appeared in the literature in the Kochen-Specker theorem \cite{ks67}, and only recently it has been preliminary employed to make statements about generalized probabilistic theories (GPTs)~\cite{hardy02,barret07,janotta22} other than quantum theory. Generalisations of this notion have been developed in recent years, and Spekkens introduced an alternative approach that extends the notion of contextuality to encompass preparations, transformations, and unsharp measurements \cite{spekkens05}, contrasting with the Kochen-Specker theorem that primarily focuses on projective measurements. Since then, generalized contextuality has been proved to subsume or be related to many common notions of nonclassicality~\cite{spekkens08, schmid20, baldi21, schmid22, liang11, schmid18, vicky22, kunjwal19, catani23, schmid2024KD, lin24}, while challenging the nonclassical status of some phenomena~\cite{catani21,INCOMPATIBILITY, TOYTHEORY,TOY2,TOY3}. Within the generalized contextuality framework -- and therefore throughout this manuscript -- a theory or scenario therein is deemed classical when it can be explained by a generalized noncontextual ontological model.
	
	Proving whether a general operational scenario is classical is not straightforward, but Ref.~\cite{selby24} presents a linear program for testing this in any arbitrary prepare-and-measure scenario and a ready-to-use implementation available in Mathematica and Python\footnote{Available in Mathematica at \\ \href{https://github.com/eliewolfe/SimplexEmbedding}{\texttt{https://github.com/eliewolfe/SimplexEmbedding}} and in Python at  \href{https://github.com/pjcavalcanti/SimplexEmbeddingGPT}{\texttt{https://github.com/pjcavalcanti/SimplexEmbeddingGPT}}.}. Formally, to check for the existence of such a noncontextual model the code instead checks for the condition of \emph{simplex- embeddability}, an equivalent notion of nonclassicality devised for GPTs~\cite{simplex}. Moreover, the code estimates how much depolarising noise must act on the experiment so that it admits of a noncontextual explanation, and it can be modified to include other noise models since it is open-source. This novel tool has already been employed to explore prepare-and-measure scenarios related to the quantum minimum-error state-discrimination protocol~\cite{rossi22} and to provide a certification of contextuality in an experimental implementation of the quantum interrogation task~\cite{giordani23}.
	
	Since we have good and operationally motivated measures for contextuality (see Sec.~\ref{sec3} for a discussion on this) -- robustness against different kinds of noise -- and we know that contextuality powers up the advantage behind parity-oblivious multiplexing tasks, it is natural to ask how these contextuality measures relate to the natural quantifier in POM tasks, i.e., the success probability itself. In this work, we turn to this problem by exploring the generality allowed by the code in Ref.~\cite{selby24} to define different kinds of noise and compare its behaviour to the success probability in POM tasks. We find a general expression relating the robustness of contextuality to depolarisation and the success rate for $n$-to-1 POM tasks and show how it can be used to recover a well-known bound on the number of bits optimally encoded in a qubit in this task. Then, building up on Ref.~\cite{rossi22}, we generalise its modification to estimate robustness of contextuality to dephasing noise in a family of scenarios related to the 3-to-1 POM. Our numerical results show that robustness of contextuality to dephasing can also perform as a quantifier when minimised over specific axes. Moreover, since the code essentially tests for simplex-embeddability of noisy states and effects, the method we adopt here can also be used for testing a class of generalized probabilistic theories broader than just quantum theory.
	
	\section{Preliminary concepts}\label{sec1}
	
	\subsection{Noncontextuality}
	The so-called operational approach is a way to construct physical theories based on laboratory experience~\cite{hardy02,barret07,dariano2017}. This means that instead of the physical properties of a system and their time evolution, operational theories focus on how this physical system can be prepared, manipulated, and investigated in general experimental setups. Therefore, this approach is broad in the sense that it can be used to describe different kinds of theories, quantum and classical theories included. A typical \emph{prepare-and-measure scenario} is described in this framework by the possible preparation procedures $P\in \mathcal{P}$, measurement procedures $M\in \mathcal{M}$, and possible outcomes $k\in K$ for each measurement that can be implemented in the experiment, as well as the statistics $p(k|M, P)$ obtained by implementing these procedures. 
	
	Moreover, these scenarios are also characterised by equivalences between different procedures: two preparations, denoted as $P$ and $P'$, are deemed \emph{operationally equivalent}, denoted as $P\simeq P'$, if they cannot be distinguished empirically, even in principle, i.e.,
	\begin{equation}
		P\simeq P'\iff p(k|M,P)=p(k|M,P'),\quad\forall k|M\in K\times\mathcal{M}.
	\end{equation}
	Similarly, two measurement outcomes, denoted as $k|M$ and $k'|M'$, are considered operationally equivalent, denoted as $k|M\simeq k'|M'$, if they both yield the same probability for any possible preparation, i.e.,
	\begin{equation}
		k|M\simeq k'|M'\iff p(k|M,P)=p(k'|M',P),\quad\forall P\in\mathcal{P}.
	\end{equation}
	Because coarse-graining over measurement outcomes is an operationally valid measurement outcome, the notion of equivalent measurement procedures follows naturally from the equivalence between measurement outcomes. Notice that the equivalence relation $\simeq$ yields \emph{equivalence classes} between the procedures, so that 
	\begin{equation}
		[P]:=\{P'\in\mathcal{P}|P'\simeq P\};
	\end{equation}
	\begin{equation}
		[k|M]:=\{k'|M'\in K\times\mathcal{M}|k'|M'\simeq k|M\}.
	\end{equation}
	The immediate interpretation is that the probability rule $p$ in the operational theory does not depend on the procedures $P$ and $k|M$ themselves, but only on the equivalence classes $[P]$ and $[k|M]$ they belong to. The excess information that singles out a particular procedure $P$ in the class $[P]$ (or a particular procedure $k|M$ in the class $[k|M]$) is what we refer to as a \emph{context}. The tuple $(\mathcal{P},\mathcal{M},K,p,\simeq)$ defines an \emph{operational theory}. As previously mentioned, quantum theory can be reformulated as one possible operational theory.
	
	Generalized contextuality (hereon referred to as \textit{contextuality} since it is the relevant notion of contextuality for the purpose of this manuscript) arises within this operational framework as the impossibility of an operational theory to conform with a noncontextual ontological model (as we describe next). An ontological model for an operational theory provides an underlying explanation for its statistics based on classical probability theory and Boolean logic, such that the behaviours captured by the operational theory can be reasoned about in classical terms. In particular, the ontological model maps the system to an \textit{ontic space} $\Lambda$, the elements $\lambda \in \Lambda$ are called \textit{ontic states}, preparation procedures are associated with probability distributions $\mu_P(\lambda)$, called \emph{epistemic states}, 
	and measurement outcomes are associated with \emph{response functions}\footnote{i.e., $\sum_{k\in K}\xi_{k|M}(\lambda)=1,\quad\forall M\in\mathcal{M},\lambda\in\Lambda.$} $\xi_{k|M}(\lambda)$  over the ontic space, with $\lambda\in\Lambda$. This model explains the operational theory when, for all $P\in\mathcal{P}$ and $k|M\in K\times\mathcal{M}$, it satisfies
	\begin{equation}
		p(k|M, P)=\sum_{\lambda\in \Lambda} \xi_{k|M}(\lambda) \mu_P(\lambda).
	\end{equation}
	
	Admitting of some ontological model is not yet a classical feature but a general one. Indeed, classicality has additional demands: any equivalences between operational procedures must hold for their representation using ontological models\footnote{Otherwise, further arguments must be provided on why the procedures are not operationally distinguishable, despite being ontologically different.}, which underpins the notion of noncontextuality~\cite{spekkens05,spekkens19lebniz}. An ontological model satisfies the assumption of noncontextuality if, for any $P, P'\in\mathcal{P}$, $k|M,k'|M'\in K\times\mathcal{M}$,
	\begin{equation}
		P \simeq P' \implies \mu_P(\lambda) = \mu_{P'}(\lambda),\quad \forall \lambda\in\Lambda,
	\end{equation}
	\begin{equation}
		k|M \simeq k'|M' \implies  \xi_{k|M}(\lambda) = \xi_{k'|M'}(\lambda),\quad  \forall\lambda\in\Lambda.
	\end{equation}
	In other words, an ontological model is noncontextual if  the contexts are not relevant for the ontological explanation underlying the experiment whenever the statistics of the operational theory do not depend on those contexts. An ontological model satisfying this constraint is said noncontextual, and a theory is contextual when it does not admit such a description.
	
	The condition of admitting of some noncontextual ontological model has proven to be a precise definition of classicality (when classicality is understood as the theory admiting of an embedding into a simplicial theory~\cite{simplex}) since the noncontextual ontological model itself provides a classical explanation to the empirical predictions. Hence whenever we say \emph{classical} throughout this article, we mean \emph{generalized-noncontextual}. As was shown before, quantum theory does not admit such explanations~\cite{spekkens05}, featuring nonclassicality.

	\subsection{Operational measures of contextuality}
	
\begin{figure}[h!]
		\centering
		\begin{tabular}{cc}
			\includegraphics{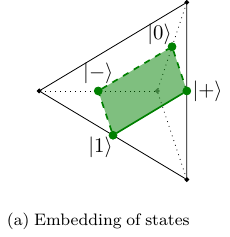} & 
			\includegraphics{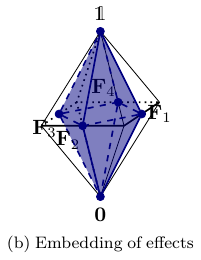}
		\end{tabular}
		\caption{\textbf{Example of simplex embedding.} In the figure we see a qualitative example of a simplex-embedding: consider a scenario with the four pure qubit states in the equator of the Bloch ball depicted in (a) (and their convex combinations) and consider the effects to be the projection onto the same states -- represented as $\{F_i\}_{i=1}^4$ in (b) -- together with the $0$ and $\mathbb{1}$ elements of POVM (and convex combinations thereof). This fragment can be represented as the green square in (a) and blue diamond-shaped object in (b). One can see that these objects can be embedded (colloquially, `fit inside') into a classical GPT: the green set of states is embedded into a four-dimensional simplex (the tetrahedron in (a)) and the effects are embedded into the dual of the tetrahedron (the cube in (b)). If this is not possible to do -- for simplices in any dimension -- then there is no simplex embedding. For any GPT that admits no simplex-embedding, one can see that by shrinking either the effects or the states via some noise process, one will always make this procedure possible.}
		\label{fig:SimplexEmbedding}
	\end{figure}

	Often, assessing noncontextuality of an operational scenario requires the characterisation of a large ontic space\footnote{In quantum scenarios, for instance, the ontic space might be as large as $|\Lambda|=(\text{dim}\mathcal{H})^2$~\cite{selby24}.} and the possible epistemic states and response functions over it that can explain the statistical data and conform to the assumption of noncontextuality for all equivalence relations. To avoid these issues when assessing the nonclassicality of the examples investigated in this work, we will employ the linear program introduced in Ref.~\cite{selby24}, which relies on results from the framework of generalized probabilistic theories (GPT)~\cite{hardy02,barret07,janotta22}. We will not enter into the details of how the operational scenario translates to the GPT description, but a summary of the linear program and the reasoning behind it is provided in Appedix~\ref{app1}. The relevant aspect for the present work is that this linear program decides the existence of a simplex embedding for the GPT fragment (see Fig.~\ref{fig:SimplexEmbedding}), which in turn is equivalent to the existence of a noncontextual ontological model for the associated operational scenario~\cite{simplex}. If the program cannot find a simplex embedding for the input scenario, i.e., the scenario exhibits contextuality, it calculates how much depolarising noise $r$ is necessary so that the simplex embedding becomes possible. The implementation of the program takes in sets of states and effects of a GPT and outputs this minimal value of $r$ along with the respective noncontextual ontological model explaining the scenario partially depolarised by a factor $r$. 
	
	The quantity $r$ hence may serve as an operational measure of nonclassicality, which we denote as the \emph{robustness of contextuality to depolarisation}. It is known from simplex embedding reasoning that there is always a finite amount of partial depolarising noise under which an operational scenario becomes noncontextual, and explicit examples of this fact have been reported~\cite{schmid18,marvian20,kunjwal15}. Alternatively, robustness can be defined in terms of different kinds of noise besides depolarising, so in this work, we also consider the case of dephasing. This is because it is a well-established fact that a completely dephased prepare-and-measure scenario will always admit of a noncontextual model~\cite{rossi22}, so the action of dephasing noise over a scenario will also eventually allow for a simplex embedding\footnote{Although in this case, partial dephasing noise might not be enough to allow for the simplex embedding, as shown in Ref.~\cite{rossi22}.}. Therefore, we have further modified the code to estimate the robustness of contextuality to dephasing noise. Our modification works for any strongly self-dual GPT with finite dimension, and the reasoning behind it is given in detail in Appendix~\ref{appendix2}. This generality has the purpose of providing a tool that can be used for research both in quantum and generalised theories, while the current work provides an application of this tool for the quantum GPT. Therefore no beyond-quantum behaviours will be explored in the coming sections.
	\FloatBarrier
	\subsection{Parity-oblivious multiplexing}\label{sec1-3}
	
	In a $n$-to-1 parity-oblivious multiplexing (POM) task, a bitstring $x$ of $n$ classical bits is encoded in the preparation $P_x$ of a quantum state. Measurements are constructed such that a binary measurement $M_y$ returns outcome $k=0$ when $x_y=0$, where $x_y$ is the $y$-th entry of the string $x$, and $k=1$ otherwise. Moreover, the encoding must be done so that no information about the parity of any subset of the string with more than one bit can be recovered from a single measurement. 
	
	Mathematically, one can define the parity of subsets of $x$ as follows: consider the parity bitstrings $\{t|t=(t_1,...,t_n),\quad \sum_it_i\geq2\}$; then, $x\cdot t$ tell us the parity of some subset of $x$ determined by $t$ -- and the condition $\sum t_i\geq 2$ ensures this subset has at least two bits. Therefore, the set $\{x\cdot t:=\sum_i x_it_i\}_t$ encodes the parity information of all subsets of $x$ containing more than one bit.
	With this notation, we say that a scenario satisfies parity-obliviousness if, given the parity strings $\{t\}$, it satisfies
	\begin{equation}
		\sum_{x|x\cdot t=0}p(k=x_y|M_y,P_x)=\sum_{x|x\cdot t=1}p(k=x_y|M_y,P_x), 
		\label{eq:parityobliviousness}
	\end{equation}
	for all $t$, $x$ and $y$. This is why generalized contextuality is of key interest for this protocol: for the $n$-to-1 POM there are $2^n$ possible preparations that must all be indistinguishable once parity about the encoded bitstring is lost. Kochen-Specker contextuality~\cite{ks67}, for instance, cannot give an account of scenarios with multiple preparations and could only assess the nonclassicality related to the task of retriving bits of a specified encoded bitstring, something that displays no quantum advantage according to Holevo's result.
	
\begin{figure}[!t]
		\includegraphics[scale=0.75]{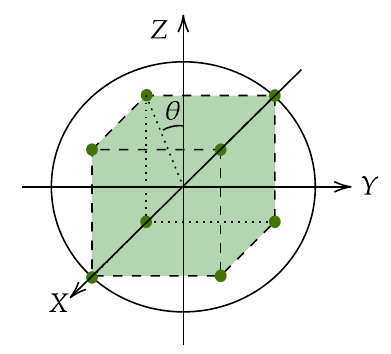}
		\centering
		\caption{ Representation of the preparation procedures of a 3-to-1 POM scenario among the family of scenarios investigated. Measurements lie on the $X$, $Y$, and $Z$ axes, while preparations are represented by the vertices of the hexahedron, lying on thee surface of the Bloch sphere. Each of the 8 possible bitstrings is encoded in a different vertex of the polytope, and fixing a value for the angle $\theta$ between all states and the $Z$ axis singles out a POM scenario.}
		\label{fig1}
	\end{figure}

	In the quantum realisation of a $3$-to-1 parity-oblivious task that is going to be explored in this work, a 3-bitstring is encoded in different preparations of a qubit.  Parity-obliviousness imposes that mixing preparations with equal parity must yield the same statistical mixture as mixing preparations with the opposite parity. This adds symmetries to the shape formed by the convex hull formed by the preparations. We exploit these symmetries by considering implementations whose set of states is given by a rectangular cuboid, with each vertex parameterized by the angle $\theta$ that all preparations equally form with respect to the $Z$ axis of the Bloch sphere. Measurements are assigned as the tomographic complete set $\hat{X},\hat{Y},\hat{Z}$ of Pauli measurements. This realisation obeys parity obliviousness constraints given in Eq.~\eqref{eq:parityobliviousness}, and a schematic representation of this scenario can be found in Figure~\ref{fig1}, where the possible states are given explicitly by
	\begin{eqnarray}\label{eq:states}
		\rho_{(x_1,x_2,x_3)}&=&\frac12\left(\mathbb{1}+\frac{(-1)^{x_1}}{\sqrt2}\sin\theta\hat{X}+\frac{(-1)^{x_2}}{\sqrt2}\sin\theta\hat{Y}\right.\nonumber\\&&\qquad\qquad\qquad\qquad\left.+(-1)^{x_3}\cos\theta\hat{Z}\right),
	\end{eqnarray}
	for $\theta\in\left[0,\frac{\pi}{2}\right]$, while the measurement outcomes are projections onto the eigenvectors of the Pauli matrices, i.e.,
	\begin{equation}\label{eq:effects}
		E_{k|M_{y}}:=\frac12(\mathbb{1}+(-1)^k\hat{M}_y),
	\end{equation}
	where $k\in\{0,1\}$ and $\hat{M}_1:=\hat{X}$, $\hat{M}_2:=\hat{Y}$, and $\hat{M}_3:=\hat{Z}$ are the Pauli operators.
	
	Notice that even though Bob cannot retrieve more than $1$ bit of information \emph{in a single-shot}, he can \emph{choose} which information to retrieve (though not deterministically). This can only be done because, even though one can at most retrieve $1$ bit of information from a qubit, the qubit can carry more information than that. In this sense, Holevo's theorem is obeyed, but communication advantage is also achieved.
	
	In general, the resourcefulness of a POM task is quantified by the probability of getting the outcome for measurement $y$ that correctly matches the $y$-th bit in the string $x$, for any outcome and string, i.e.,
	\begin{equation}
		s = \frac{1}{2^n n}\sum^n_{ y=1}\,\sum_{x\in X_n}p(k=x_y|M_y,P_x),
	\end{equation}
	where $X_n:=\{0,1\}^{\times n}$ is the set of all possible bitstrings that can be encoded in the task. It has been demonstrated that indeed the success rate is a good quantifier for a resource theory of contextuality~\cite{catani24}. Moreover, it is known that quantum realisations of this protocol can exceed the noncontextual bound of $s_{NC}=\frac12\left(1+\frac{1}{n}\right)$~\cite{spekkens09}, attesting that contextuality is the source of quantum advantage for these tasks, since quantum realisations can achieve success rates of up to $s_Q=\frac12\left(1+\frac{1}{\sqrt{n}}\right)$~\cite{chailloux16}, constituting instances of quantum advantage for communication.\footnote{In fact, only the equivalence classes for preparations are used to prove the quantum advantage for POM tasks in Ref.~\cite{spekkens09}, leading to the claim that preparation contextuality is the source of this advantage. However, given that the only equivalence class for measurement outcomes in this scenario is the trivial one, there is no difference between generalised and preparation noncontextuality in this case.}
	\FloatBarrier
	\section{Results}
	\subsection{Robustness to depolarising noise in $n$-to-1 parity-oblivious multiplexing}\label{sec2}
	
	Given that the advantage in POM tasks is underpinned by contextuality, and that depolarizing noise may transform a contextual scenario into a noncontextual one, it is natural to ask how much noise a POM scenario must receive to perform as poorly as classical realisations thereof. In particular, in the case of partial depolarising noise in a quantum realisation, every state under the action of this noise will be described by the mixture
	\begin{equation}
		\rho_x\mapsto(1-r)\rho_x+r\mu,
	\end{equation}
	where $\mu$ is the maximally mixed state, and $r\in[0,1]$ is the noise quantifier. Notice that this represents a modification in all probabilities obtained in the scenario, since
	\begin{eqnarray}
		p(k=x_y|P_x,M_y) &=& \Tr(E_{k=x_y|M_y}\rho_x)\\
		&\mapsto& (1-r)\Tr(E_{k=x_y|M_y}\rho_x)\nonumber\\
		&&\qquad\qquad\quad+r\Tr(E_{k=x_y|M_y}\mu) \\
		&=&(1-r)p(k=x_y|P_x,M_y)\nonumber\\
		&&\qquad\qquad\quad+ r\Tr(E_{k=x_y|M_y}\mu),
	\end{eqnarray}
	where $E_{k=x_y|M_y}$ are the POVM elements of measurement $M_y$. Once we calculate the success rate for the depolarised version of the experiment, we find
	
	\begin{align}
		s_{depol} &= \frac{(1-r)}{2^nn}\sum_{y=1}^n\sum_{x\in[0,1]^{\times n}}p(k=x_y|P_x,M_y)\nonumber\\ &\qquad\qquad+\frac{r}{2^nn}\sum_{y=1}^n\sum_{x\in X_n}\Tr(E_{k=x_y|M_y}\mu)\\
		&= (1-r)s +\frac{r}{2n}\sum_{y=1}^n\Tr(\mu)\\
		&= (1-r)s+\frac{r}{2},
		\label{eq:sdepol}
	\end{align}
	where the second equality follows from the fact that for every choice of measurement $y$, there will be $2^{n-1}$ strings $x$ with $y$-th input equal to $0$, and $2^{n-1}$ strings with the same input equal to $1$. Since all $E_{k=x_y|M_y}$ are elements of POVM, it must be that $E_{k=0_y|M_y}+E_{k=1_y|M_y}=\mathbb{1}$ for any $M_y$, simplifying the summation. 
	
	Let us define the \emph{robustness of contextuality to depolarisation}, $r^{depol}_{min}$, as the minimum partial depolarising noise to be added to a scenario such that its depolarised version performs as poorly as a noncontextual realisation. In other words, $r_{min}^{depol}$ is the minimum depolarizing noise such that the success rate of the depolarised scenario is the optimal noncontextual one, $s_{NC}$. Substituting this condition into Eq.~\eqref{eq:sdepol}, we get
	\begin{equation}\label{robustness1}
		r_{min}^{depol}=\frac{s-s_{NC}}{s-\frac12}.
	\end{equation}
	This proves the following result:
	
	\begin{proposition}\label{prop1}
		In any $n$-to-1 parity-oblivious multiplexing scenario, success rate and robustness of contextuality to depolarisation are monotonically equivalent resource quantifiers via Eq.~\eqref{robustness1}.
	\end{proposition}
	Proposition~\ref{prop1} then links a geometrical feature of states and effects used to implement a POM task and their success in such a task:  the higher the depolarizing noise needed to make states $P_x$ and measurements $M_y$ become simplex-embeddable, the best they perform in the POM task, and vice-versa.
	
	With this relation, one can derive bounds on the number of bits that can be optimally encoded in a particular quantum system by assuming bounds on the maximal robustness to depolarising achievable for this system. For instance, for qubits, it is common to assume that $r_{min}^{depol}$ can never exceed the value $\frac12$ for whatever signature of quantum advantage taken into consideration, in particular $r_{min}^{depol}<\frac12$ for a finite amount of extremal states and effects in the fragment (which is always the case in $n$-to$-1$ POM tasks with finite $n$). We provide a compelling numerical induction for this bound in Appendix~\ref{appendix3} that relies on the simplex embeddability of an ever-increasing set of preparations and measurements. Then notice that the optimal quantum success rate is given by $s=\frac12(1+\frac{1}{\sqrt{n}})$~\cite{chailloux16}, while $s_{NC}=\frac12(1+\frac{1}{n})$~\cite{spekkens08}. If $r_{min}^{depol}<\frac12$, then
	
	\begin{eqnarray}
		\frac12&>&\frac{s-s_{NC}}{s-\frac12} \\
		&=& \frac{\frac{1}{\sqrt{n}}-\frac{1}{n}}{\frac{1}{\sqrt{n}}}\\
		&=&1-\frac{1}{\sqrt{n}},
	\end{eqnarray}
	which means that $n<4$. In other words:
	
	\begin{proposition} 
		\label{proposition:boundOnn}A two-levels quantum system used in a $n$-to-$1$ POM task can achieve a maximal advantage over classical systems only if $n<4$.
	\end{proposition}
	
	This result has been derived before by geometrical~\cite{hayashi06} and locality~\cite{ghorai18} arguments, but Proposition~\ref{prop1} allows us to reframe it solely on simplicial embedding assumptions. This method can, in principle, be generalized: if one finds bounds on robustness for quantum systems of higher dimensions, one can immediately use Eq.~\eqref{robustness1} to derive an upper bound to the number of optimally encoded bits. Notice that this bound holds exclusively for quantum thory, and there are GPTs other than quantum that are known to achieve a higher   success rate (including the algebraic maximum) for this protocol~\cite{banik15}.
	
	The linear program from Ref.~\cite{selby24} can also be used to get numerical results for different quantum implementations of an $n$-to-1 POM task. We focus on the case of a qubit and, given Prop.~\ref{proposition:boundOnn}, the most interesting cases are $n=2$ or $n=3$. We will, however, focus on POM scenarios encoding 3 bits in a qubit, as introduced in Sec.~\ref{sec1-3}. We deem these examples as more engaging than the simplest 2-to-1 POM since the latter is equivalent to the simplest prepare-and-measure scenario for which contextuality can be demonstrated, and much is already known about it~\cite{pusey18,khoshbin24,catani22,catani23-2,catani23}. 
	
	\begin{figure}[h!]
		\centering
		\begin{tabular}{cc}
			\includegraphics[scale=0.3]{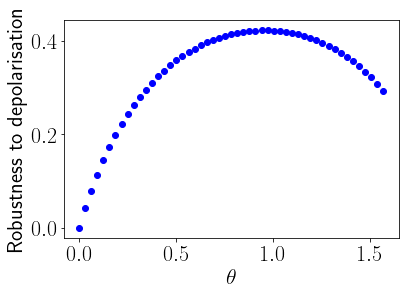} &
			\includegraphics[scale=0.3]{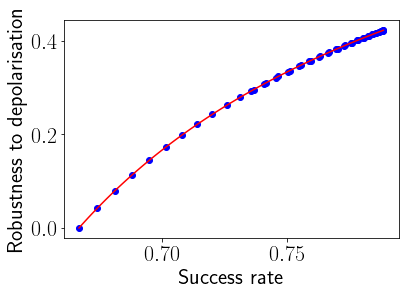}
		\end{tabular}
		\caption{Robustness of contextuality to depolarising noise vs. parameter $\theta$ and vs. success rate for the 3-to-1 POM task. The red curve represents the plot of Eq.~\ref{robustness1} for the 3-to-1 task.}
		\label{fig2}
	\end{figure}
	
	In order to showcase this relation between robustness to depolarisation and success rate in a concrete example, we input the family of scenarios introduced in Sec.~\ref{sec1-3} to the linear program, obtaining the plots in Figure~\ref{fig2}. Notice that, as dictated by Prop.~\ref{prop1}, robustness of contextuality to depolarising noise grows monotonically with the success rate. In particular, when $n=3$, we have $s_{NC}=\frac23$, and the numerical plot reproduces Eq.~\eqref{robustness1}. We can also see that the peak on robustness of contextuality is achieved when $\theta=\frac{\pi}{3}$, i.e. when the states form a perfect cube inside the Bloch sphere. This scenario is compatible with the case in Ref.~\cite{spekkens09} for which the success rate is optimal. Intuitively, this can be related to the fact that this set of states has the greatest volume among the family of scenarios, and this situation would be expected to need the most noise for a simplex embedding to exist.
	
	\subsection{Robustness to dephasing noise in 3-to-1 parity oblivious multiplexing}\label{sec3}
	
	Robustness-based quantifiers are advantageous in general since there is a fair knowledge about their properties, for instance, that in many resource theories, they are monotonic under linear combinations of free operations~\cite{barb18}. Moreover, robustness has a strong operational appeal since it relates to the ubiquitous characteristic in laboratories of uncontrolled degrees of freedom, and hence it would be interesting if statements about robustness of contextuality to other types of noise could be produced for the parity-oblivious tasks. Dephasing noise, for instance, is universally present in quantum computation and represents a considerable obstacle to the implementation of many protocols, especially concerning scalability. Moreover, it displays similarities with depolarising noise, in the sense that complete dephasing will deem any prepare-and-measure scenario noncontextual~\cite{rossi22}. In this section, we thus analyze robustness to dephasing and how this contextuality measure relates to quantifying advantage in POM tasks. However, general results such as Prop.~\ref{prop1} are challenging when it comes to dephasing since it is a more structured type of noise, making arguments of this type difficult to build. We will therefore rely on the numerical investigation of the proposed 3-to-1 POM scenarios to investigate the usefulness of robustness to dephasing in quantifying the nonclassical advantage for these tasks by employing the modified code (see Appendix~\ref{appendix2}).
	
\begin{figure}[h]
		\centering
		\begin{tabular}{cc}
			\includegraphics[scale=0.28]{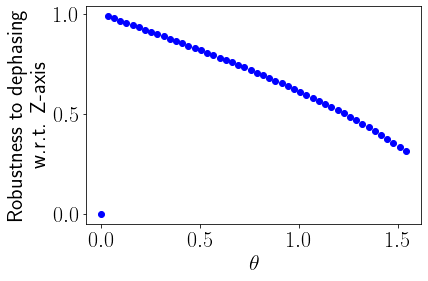} &
			\includegraphics[scale=0.28]{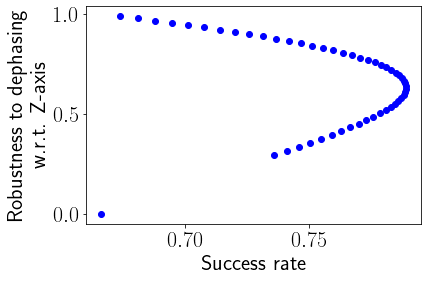}
		\end{tabular}
		\caption{Robustness of contextuality to dephasing noise vs. parameter $\theta$ and vs. success rate for the 3-to-1 POM task.}
		\label{fig3}
	\end{figure}

	The following results were obtained by inputting the states and effects from Eq.~\eqref{eq:states} and ~\eqref{eq:effects}, respectively, to the modified linear program (Appendix~\ref{appendix2}). For practical purposes, we chose the $\hat{Z}$ axis as the preserved axis, although the symmetry of this scenario should imply similar results for the $\hat{X}$ and $\hat{Y}$ axes. The results of our numerical explorations are presented in Fig.~\ref{fig3}. As was previously shown in Ref.~\cite{rossi22}, the quantum advantage gets increasingly robust to dephasing as the preparations overlap with the dephasing basis. This however is not a good quantifier of this quantum advantage, since for the particular case in which robustness of contextuality is maximal, the success rate is very close to the classical bound, and robustness of contextuality decreases for scenarios in which the success rate improves. Since achieving a success rate $s>s_{NC}$ is the only contextuality inequality in this scenario, we can safely deem robustness to dephasing with respect to the $\hat{Z}$ axis as a bad quantifier\footnote{The reason why this is the unique contextuality inequality is that $r>0$ (be it robustness to depolarising or dephasing) happens if and only if there is no noncontextual ontological model for the scenario, and from Prop.~\ref{prop1} this is equivalent to having $s>s_{NC}$.}. It is interesting to see that, differently from the depolarising case, there are scenarios achieving a better than classical success rate that can endure a noise larger than $0.5$.
	
	Notice that the statement that this is not a good quantifier tells about how well the value of $r$ captures the resourcefulness of the scenario for this particular task in contrast to a classical realisation. However, robustness, in this case, is still a good \emph{certifier} of nonclassicality, since as long as it is nonzero, it attests to the impossibility of a noncontextual ontological model for the scenario. This is possibly related to the axial symmetry of dephasing noise, when compared to the radial symmetry of the scenarios under investigation. Dephasing noise with respect to a particular axis is therefore not in a good fit with the symmetry suggested by the task, which motivates our following result.
	
	We further show that robustness to dephasing can still perform as a good quantifier by exploiting the symmetry of this scenario and minimizing the robustness of contextuality to dephasing over all relevant dephasing bases. In particular, we here consider all available measurement basis $\hat{X},\hat{Y},\hat{Z}$. To gain intuition on why, notice that these axes will maximise the action of the noise over the volume of the set of states, requiring a smaller amount of noise to produce a greater change. Allowing for axes that align with a particular preparation, for instance, would possibly yield a larger value of robustness since the polyhedron from Fig.~\ref{fig1} would keep one of its diagonals unchanged, and the value of robustness with respect to this axis would be discarded by the minimisation. Indeed, we show the numerical investigation leading to this choice in Appendix~\ref{app4}. 
	
	\begin{figure}[h]
		\centering
		\begin{tabular}{cc}
			\includegraphics[scale=0.28]{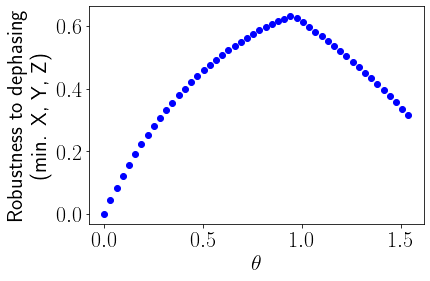} &
			\includegraphics[scale=0.28]{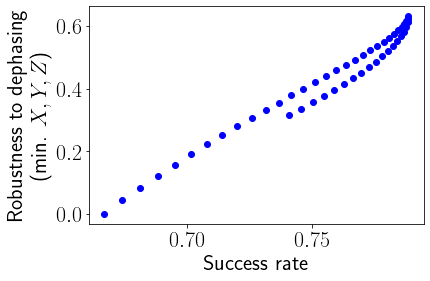}
		\end{tabular}
		\caption{Minimal robustness of contextuality to dephasing noise across all of Bob's measurements vs. parameter $\theta$ and vs. success rate for the 3-to-1 POM task.}
		\label{fig4}
	\end{figure}
	
	The plots are given in Figure~\ref{fig4}. Maximal robustness of contextuality is now not achieved for small values of $\theta$ since, for the same scenarios, there is another dephasing axis with respect to which robustness of contextuality is almost null. In particular, dephasing with respect to either $\hat{X}$ or $\hat{Y}$ axes is the minimal one up until the parameter $\theta$ reaches $\theta=\frac{\pi}{3}$, precisely the scenario to which success rate is maximal. From that point on, the minimal robustness to dephasing is the one with respect to the $\hat{Z}$ axis.
	
	The reason why this is still a good quantifier is that the extremal cases in which contextuality (and therefore success rate) is minimal and maximal are captured as extremal cases by the minimised robustness to dephasing, i.e., it assigns $r=0$ only for $s=s_N$, and its highest value coincides only with $s=s_Q$. Evidently, this minimised robustness distinguishes scenarios that the success rate does not and vice-versa, so we cannot deem them as monotonically proportional. However, specifying a value of $r$ in most cases singles out the value of $s$, and in the worst scenario it is upper and lower bounded by the success rate. These bounds do increase monotonically when they exist, which displays some sort of coherence between these two quantifiers~\cite{regula17}.
	
	It is interesting to see how this quantity displays a coherent relation with the success rate similar to the depolarising case, even though robustness of contextuality is not minimised over all possible axes in the Bloch sphere. It would be interesting to investigate whether this quantifier, when taken with respect to all possible axes, has any relation to quantities that are Bargmann invariants such as basis-independent coherence~\cite{galvao20,wagner22A}.
	
	\section{Closing remarks}\label{sec4}
	
	Our work builds up on the results from Refs.~\cite{selby24,rossi22} to further explore the role of different types of noise over the possibility of a noncontextual ontological model in prepare-and-measure scenarios. We derive an analytical relation between robustness of contextuality to depolarisation and the success rate that usually quantifies the quantum advantage in POM tasks (Prop.~\ref{prop1}). We then use it to recover a result that has been previously derived from both geometrical~\cite{hayashi06} and locality arguments~\cite{ghorai18} that a POM with optimal success rate is impossible when encoding more than 3 bits in a qubit. In particular, we give a re-framing of these proofs solely on simplex-embedding arguments, and we believe that more general statements can be made by observing the relation between robustness of contextuality and success rate, increasing the insight provided by these proofs. Our derivation contrasts with the previous ones because (1) the geometrical argument~\cite{hayashi06} is only valid for 2-levels systems since it relies on the fact that the valid set of states for these systems is spherical, while our derivation can be extended   for higher-dimensional systems; and (2) the nonlocality argument~\cite{ghorai18} relies on mapping the data from this task to a Bell scenario, which is causally distinct from the original one, while Proposition~\ref{proposition:boundOnn} is grounded on the same causal structure that generated the data.
	
	Furthermore, we present a generalisation of the linear program from Ref.~\cite{selby24} estimating robustness of contextuality to dephasing noise for any GPT fragment of a strongly self-dual GPT, with quantum theory as the example of interest, and leverage it to explore the particular case of the 3-to-1 parity-oblivious multiplexing task, to which contextuality is known to be a resource. We conclude that beyond robustness of contextuality to depolarisation, other noise models (such as a minimisation of robustness of contextuality to dephasing over a set of bases) can also display a consistent relation with respect to the standard resource quantifier. In contrast, robustness to dephasing with respect to the $\hat{Z}$ axis does not display such a relation. It however remains a good certifier of nonclassical advantage, since for any nonzero value of robustness, advantage over classical implementations is ensured. These results showcase the versatility of the linear program introduced in Ref.~\cite{selby24} since many experimental constraints can be taken into consideration when deciding the nonclassicality of a prepare-and-measure scenario.
	
	It would be interesting to investigate how the quantifiers explored in our framework relate to quantifiers of Bargmann invariants~\cite{galvao20, wagner22A} and quantifiers obtained by SDP hierarchies~\cite{chaturvedi21,chaturvedi24}.
	
	Another direction of future research includes a better understanding of how noise models beyond depolarisation should be defined in more complex GPTs, for instance, in weakly self-dual GPTs. The operational approach to quantum resource theories has gained space in the literature~\cite{selby20,bruscemi20,zjawin23,catani24} and might provide useful insights on the relation between different signatures of nonclassicality.\\
	
	\section*{Acknowledgements}
	We thank Bárbara Amaral for all the support and guidance on this manuscript, and Rafael Wagner, Giulio Camillo, and Anubhav Chaturvedi for the fruitful discussion. A.M.F. was supported by the São Paulo Research Foundation (FAPESP)  via Project Numbers 2022/05455-5 (FAS-Masters) and 2022/13208-8 (RIA-Research Internship Abroad), and from the Instituto Serrapilheira, Grant 2021/4. V.P.R. was supported by the Foundation for Polish Science (IRAP project, ICTQT, contract no. MAB/2018/5, co-financed by EU within Smart Growth Operational Programme). J.H.S. was supported by the National Science Centre, Poland (Opus project, Categorical Foundations of the Non-Classicality of Nature, Project
	No. 2021/41/B/ST2/03149). V.P.R., R.D.B., and A.B.S. acknowledge support by the Digital Horizon Europe project FoQaCiA, Foundations of quantum computational advantage, GA No. 101070558, funded by the European Union, NSERC (Canada), and UKRI
	(UK). Some figures were prepared using TikZit and Mathcha. This research was partially conducted while visiting the Okinawa Institute of Science and Technology (OIST) through the Theoretical Sciences Visiting Program (TSVP).
	
	\bigskip
	\bigskip
	\bibliographystyle{apsrev4-1}

\begin{thebibliography}{}
		\expandafter\ifx\csname natexlab\endcsname\relax\def\natexlab#1{#1}\fi
		\def\au#1{#1} \def\ed#1{#1} \def\yr#1{#1}\def\at#1{#1}\def\jt#1{{#1}}
		\def\bt#1{#1}\def\bvol#1{\textbf{#1}} \def\vol#1{#1} \def\pg#1{#1}
		\def\publ#1{#1}\def\arxiv#1{#1}\def\org#1{#1}\def\st#1{\textit{#1}}\def\series#1{#1}
		
		
		\bibitem[1]{holevo73}
		{\au{A. S. Holevo}}, \at{\textit{Bounds for the Quantity of Information Transmitted by a Quantum Communication Channel}}, \jt{Problems Inform. Transmission}, \bvol{9} (\yr{1973}) \pg{{177}--{183}}.
		
		\bibitem[2]{gisin07}
		{\au{N. Gisin} and \au{R. Thew}}, \at{\textit{Quantum communication}}, \jt{Nature Photonics}, \bvol{1}(3) (\yr{2007}) \pg{{165}--{171}}.
		
		\bibitem[3]{brassard03}
		{\au{G. Brassard}}, , \jt{Foundations of Physics}, \bvol{33}(11) (\yr{2003}) \pg{{1593}--{1616}}.
		
		\bibitem[4]{spekkens08}
		{\au{R. W. Spekkens}}, \at{\textit{``Negativity and Contextuality are Equivalent Notions of Nonclassicality''}}, \jt{Phys. Rev. Lett.}, \bvol{101} (\yr{2008}) \pg{020401}.
		
		\bibitem[5]{spekkens05}
		{\au{R. W. Spekkens}}, \at{\textit{Contextuality for preparations, transformations, and unsharp measurements}}, \jt{Phys. Rev. A}, \bvol{71} (\yr{2005}) \pg{052108}.
		
		\bibitem[6]{saha2019}
		{\au{D. Saha} and \au{A. Chaturvedi}}, \at{\textit{``Preparation contextuality as an essential feature underlying quantum communication advantage''}}, \jt{Phys. Rev. A}, \bvol{100} (\yr{2019}) \pg{022108}.
		
		\bibitem[7]{sumit23}
		{\au{A. S. S.}, \au{S. Mukherjee}, and \au{A. K. Pan}}, \at{\textit{Robust certification of unsharp instruments through sequential quantum advantages in a prepare-measure communication game}}, \jt{Phys. Rev. A}, \bvol{107} (\yr{2023}) \pg{012411}.
		
		\bibitem[8]{roch22}
		{\au{C. Roch i Carceller}, \au{K. Flatt}, \au{H. Lee}, \au{J. Bae}, and \au{J. B. Brask}}, \at{\textit{Quantum vs Noncontextual Semi-Device-Independent Randomness Certification}}, \jt{Phys. Rev. Lett.}, \bvol{129} (\yr{2022}) \pg{050501}.
		
		\bibitem[9]{schmid22}
		{\au{D. Schmid}, \au{H. Du}, \au{J. H. Selby}, and \au{M. F. Pusey}}, \at{\textit{``Uniqueness of Noncontextual Models for Stabilizer Subtheories''}}, \jt{Phys. Rev. Lett.}, \bvol{129} (\yr{2022}).
		
		\bibitem[10]{bowles23}
		{\au{J. Bowles}, \au{V. J. Wright}, \au{M. Farkas}, \au{N. Killoran}, and \au{M. Schuld}}, \at{\textit{Contextuality and inductive bias in quantum machine learning}}, \jt{arXiv:2302.01365 [quant-ph]} (\yr{2023}).
		
		\bibitem[11]{spekkens09}
		{\au{R. W. Spekkens}, \au{D. H. Buzacott}, \au{A. J. Keehn}, \au{B. Toner}, and \au{G. J. Pryde}}, \at{\textit{Preparation Contextuality Powers Parity-Oblivious Multiplexing}}, \jt{Phys. Rev. Lett.}, \bvol{102} (\yr{2009}) \pg{010401}.
		
		\bibitem[12]{ambainis19}
		{\au{A. Ambainis}, \au{M. Banik}, \au{A. Chaturvedi}, \au{D. Kravchenko}, and \au{A. Rai}}, \at{\textit{Parity Oblivious $d$-Level Random Access Codes and Class of Noncontextuality Inequalities}}, \jt{arXiv:1607.05490 [quant-ph]} (\yr{2016}).
		
		\bibitem[13]{chailloux16}
		{\au{A. Chailloux}, \au{I. Kerenidis}, \au{S. Kundu}, and \au{J. Sikora}}, \at{\textit{``Optimal bounds for parity-oblivious random access codes''}}, \jt{New J. Phys.}, \bvol{18} (\yr{2016}) \pg{045003}.
		
		\bibitem[14]{yadavalli22}
		{\au{S. A. Yadavalli} and \au{R. Kunjwal}}, \at{\textit{``Contextuality in entanglement-assisted one-shot classical communication''}}, \jt{Quantum}, \bvol{6} (\yr{2022}) \pg{839}.
		
		\bibitem[15]{lostaglio20}
		{\au{M. Lostaglio}}, \at{\textit{``Certifying Quantum Signatures in Thermodynamics and Metrology via Contextuality of Quantum Linear Response''}}, \jt{Phys. Rev. Lett.}, \bvol{125} (\yr{2020}) \pg{230603}.
		
		\bibitem[16]{lostaglio22}
		{\au{M. Lostaglio} and \au{G. Senno}}, \at{\textit{``Contextual advantage for state-dependent cloning''}}, \jt{Quantum}, \bvol{4} (\yr{2020}) \pg{258}.
		
		\bibitem[17]{schmid18}
		{\au{D. Schmid} and \au{R. W. Spekkens}}, \at{\textit{``Contextual Advantage for State Discrimination''}}, \jt{Phys. Rev. X}, \bvol{8} (\yr{2018}) \pg{011015}.
		
		\bibitem[18]{shin21}
		{\au{J. Shin}, \au{D. Ha}, and \au{Y. Kwon}}, \at{\textit{``Quantum Contextual Advantage Depending on Nonzero Prior Probabilities in State Discrimination of Mixed Qubit States''}}, \jt{Entropy}, \bvol{23} (\yr{2021}) \pg{1583}.
		
		\bibitem[19]{flatt22}
		{\au{K. Flatt}, \au{H. Lee}, \au{C. R. I. Carceller}, \au{J. B. Brask}, and \au{J. Bae}}, \at{\textit{``Contextual Advantages and Certification for Maximum-Confidence Discrimination''}}, \jt{PRX Quantum}, \bvol{3} (\yr{2022}) \pg{030337}.
		
		\bibitem[20]{mukherjee22}
		{\au{S. Mukherjee}, \au{S. Naonit}, and \au{A. K. Pan}}, \at{\textit{``Discriminating three mirror-symmetric states with a restricted contextual advantage''}}, \jt{Phys. Rev. A}, \bvol{106} (\yr{2022}) \pg{012216}.
		
		\bibitem[21]{ks67}
		{\au{S. Kochen} and \au{E. P. Specker}}, \at{\textit{The Problem of Hidden Variables in Quantum Mechanics}}.
		
		\bibitem[22]{hardy02}
		{\au{L. Hardy}}, \at{\textit{``Quantum Theory From Five Reasonable Axioms''}}, \jt{arXiv:quant-ph/0101012} (\yr{2001}).
		
		\bibitem[23]{barret07}
		{\au{J. Barrett}}, \at{\textit{``Information processing in generalized probabilistic theories''}}, \jt{Phys. Rev. A}, \bvol{75} (\yr{2007}) \pg{032304}.
		
		\bibitem[24]{janotta22}
		{\au{P. Janotta} and \au{H. Hinrichsen}}, \at{\textit{``Generalized probability theories: what determines the structure of quantum theory?''}}, \jt{J. Phys. A: Math. Theor.}, \bvol{47} (\yr{2014}) \pg{323001}.
		
		\bibitem[25]{schmid20}
		{\au{D. Schmid}, \au{J. H. Selby}, and \au{R. W. Spekkens}}, \at{\textit{``Unscrambling the omelette of causation and inference: The framework of causal-inferential theories''}}, \jt{arXiv:2009.03297 [quant-ph]} (\yr{2020}).
		
		\bibitem[26]{baldi21}
		{\au{R. D. Baldijão}, \au{R. Wagner}, \au{C. Duarte}, \au{B. Amaral}, and \au{M. T. Cunha}}, \at{\textit{``Emergence of Noncontextuality under Quantum Darwinism''}}, \jt{PRX Quantum}, \bvol{2} (\yr{2021}) \pg{030351}.
		
		\bibitem[27]{liang11}
		{\au{Y. C. Liang}, \au{R. W. Spekkens}, and \au{H. M. Wiseman}}, \at{\textit{``Specker’s parable of the overprotective seer: A road to contextuality, nonlocality and complementarity''}}, \jt{Physics Reports}, \bvol{506} (\yr{2011}) \pg{{1}--{39}}.
		
		\bibitem[28]{vicky22}
		{\au{V. Wright} and \au{M. Farkas}}, \at{\textit{``An invertible map between Bell non-local and contextuality scenarios''}}, \jt{arXiv:2211.12550 [quant-ph]} (\yr{2022}).
		
		\bibitem[29]{kunjwal19}
		{\au{R. Kunjwal}, \au{M. Lostaglio}, and \au{M. F. Pusey}}, \at{\textit{``Anomalous weak values and contextuality: Robustness, tightness, and imaginary parts''}}, \jt{Phys. Rev. A}, \bvol{100} (\yr{2019}) \pg{042116}.
		
		\bibitem[30]{catani23}
		{\au{L. Catani}, \au{M. Leifer}, \au{D. Schmid}, and \au{R. W. Spekkens}}, \at{\textit{Why interference phenomena do not capture the essence of quantum theory}}, \jt{Quantum}, \bvol{7} (\yr{2023}) \pg{1119}.
		
		\bibitem[31]{schmid2024KD}
		{\au{D. Schmid}, \au{R. D. Baldijão}, \au{Y. Yīng}, \au{R. Wagner}, and \au{J. H. Selby}}, \at{\textit{Kirkwood-Dirac representations beyond quantum states (and their relation to noncontextuality)}}, \jt{arXiv:2405.04573 [quant-ph]} (\yr{2024}).
		
		\bibitem[32]{lin24}
		{\au{J. Lin}, \au{K. Wang}, \au{L. Xiao}, \au{D. Qu}, \au{G. Zhu}, \au{Y. Zhang}, and \au{P. Xue}}, \at{\textit{Experimental investigation of contextual robustness and coherence in state discrimination}}, \jt{Phys. Rev. A}, \bvol{109} (\yr{2024}) \pg{052208}.
		
		\bibitem[33]{catani21}
		{\au{L. Catani}, \au{M. Leifer}, \au{D. Schmid}, and \au{R. W. Spekkens}}, \at{\textit{``Why interference phenomena do not capture the essence of quantum theory''}}, \jt{arXiv:2111.13727 [quant-ph]} (\yr{2021}).
		
		\bibitem[34]{INCOMPATIBILITY}
		{\au{J. H. Selby}, \au{D. Schmid}, \au{E. Wolfe}, \au{A. B. Sainz}, \au{R. Kunjwal}, and \au{R. W. Spekkens}}, \at{\textit{``Contextuality without incompatibility''}}, \jt{arXiv:2106.09045 [quant-ph]} (\yr{2021}).
		
		\bibitem[35]{TOYTHEORY}
		{\au{R. W. Spekkens}}, \at{\textit{``Evidence for the epistemic view of quantum states: A toy theory''}}, \jt{Phys. Rev. A}, \bvol{75} (\yr{2007}) \pg{032110}.
		
		\bibitem[36]{TOY2}
		{\au{S. D. Barlett}, \au{T. Rudolph}, and \au{R. W. Spekkens}}, \at{\textit{``Reconstruction of Gaussian quantum mechanics from Liouville mechanics with an epistemic restriction''}}, \jt{Phys. Rev. A}, \bvol{86} (\yr{2012}) \pg{012103}.
		
		\bibitem[37]{TOY3}
		{\au{R. W. Spekkens}}, \at{\textit{``Quasi-Quantization: Classical Statistical Theories with an Epistemic Restriction''}}, \bt{Quantum Theory: Informational Foundations and Foils}, \bvol{181}, (\yr{2016}).
		
		\bibitem[38]{selby24}
		{\au{J. H. Selby}, \au{E. Wolfe}, \au{D. Schmid}, \au{A. B. Sainz}, and \au{V. P. Rossi}}, \at{\textit{Linear Program for Testing Nonclassicality and an Open-Source Implementation}}, \jt{Phys. Rev. Lett.}, \bvol{132} (\yr{2024}) \pg{050202}.
		
		\bibitem[39]{simplex}
		{\au{D. Schmid}, \au{J. H. Selby}, \au{E. Wolfe}, \au{R. Kunjwal}, and \au{R. W. Spekkens}}, \at{\textit{Characterization of Noncontextuality in the Framework of Generalized Probabilistic Theories}}, \jt{PRX Quantum}, \bvol{2} (\yr{2021}) \pg{010331}.
		
		\bibitem[40]{rossi22}
		{\au{V. P. Rossi}, \au{D. Schmid}, \au{J. H. Selby}, and \au{A. B. Sainz}}, \at{\textit{Contextuality with vanishing coherence and maximal robustness to dephasing}}, \jt{Phys. Rev. A}, \bvol{108} (\yr{2023}) \pg{032213}.
		
		\bibitem[41]{giordani23}
		{\au{T. Giordani}, \au{R. Wagner}, \au{E. Esposito}, \au{A. Camillini}, \au{F. Hoch}, \au{G. Carvacho}, \au{C. Pentangelo}, \au{F. Ceccarelli}, \au{S. Piacentini}, \au{A. Crespi}, \au{N. Spagnolo}, \au{R. Osellame}, \au{E. F. Galvão}, and \au{F. Sciarrino}}, \at{\textit{Experimental certification of contextuality, coherence, and dimension in a programmable universal photonic processor}}, \jt{Science Advances}, \bvol{9}(44) (\yr{2023}) \pg{eadj4249}.
		
		\bibitem[42]{dariano2017}
		{\au{G. M. D’Ariano}, \au{G. Chiribella}, and \au{P. Perinotti}}, \at{\textit{Quantum Theory from First Principles: An Informational Approach}},  (\yr{2017}) \publ{Cambridge University Press}.
		
		\bibitem[43]{spekkens19lebniz}
		{\au{R. W. Spekkens}}, \at{\textit{``The ontological identity of empirical indiscernibles: Leibniz's methodological principle and its significance in the work of Einstein''}}, \jt{arXiv:1909.04628 [physics.hist-ph]} (\yr{2019}).
		
		\bibitem[44]{marvian20}
		{\au{I. Marvian}}, \at{\textit{``Inaccessible information in probabilistic models of quantum systems, non-contextuality inequalities and noise thresholds for contextuality''}}, \jt{arXiv:2003.05984 [quant-ph]} (\yr{2020}).
		
		\bibitem[45]{kunjwal15}
		{\au{R. Kunjwal} and \au{R. W. Spekkens}}, \at{\textit{``From the Kochen-Specker Theorem to Noncontextuality Inequalities without Assuming Determinism''}}, \jt{Phys. Rev. Lett.}, \bvol{115} (\yr{2015}) \pg{110403}.
		
		\bibitem[46]{catani24}
		{\au{L. Catani}, \au{T. D. Galley}, and \au{T. Gonda}}, \at{\textit{Resource-theoretic hierarchy of contextuality for general probabilistic theories}}, \jt{arXiv:2406.00717 [quant-ph]} (\yr{2024}).
		
		\bibitem[47]{hayashi06}
		{\au{M. Hayashi}, \au{K. Iwama}, \au{H. Nishimura}, \au{R. Raymond}, and \au{S. Yamashita}}, \at{\textit{(4,1)-Quantum random access coding does not exist—one qubit is not enough to recover one of four bits}}, \jt{New Journal of Physics}, \bvol{8}(8) (\yr{2006}) \pg{129}.
		
		\bibitem[48]{ghorai18}
		{\au{S. Ghorai} and \au{A. K. Pan}}, \at{\textit{Optimal quantum preparation contextuality in an $n$-bit parity-oblivious multiplexing task}}, \jt{Phys. Rev. A}, \bvol{98} (\yr{2018}) \pg{032110}.
		
		\bibitem[49]{banik15}
		{\au{M. Banik}, \au{S. S. Bhattacharya}, \au{A. Mukherjee}, \au{A. Roy}, \au{A. Ambainis}, and \au{A. Rai}}, \at{\textit{Limited preparation contextuality in quantum theory and its relation to the Cirel'son bound}}, \jt{Phys. Rev. A}, \bvol{92} (\yr{2015}) \pg{030103(R)}.
		
		\bibitem[50]{pusey18}
		{\au{M. F. Pusey}}, \at{\textit{Robust preparation noncontextuality inequalities in the simplest scenario}}, \jt{Phys. Rev. A}, \bvol{98} (\yr{2018}) \pg{022112}.
		
		\bibitem[51]{khoshbin24}
		{\au{M. Khoshbin}, \au{L. Catani}, and \au{M. Leifer}}, \at{\textit{Alternative robust ways of witnessing nonclassicality in the simplest scenario}}, \jt{Phys. Rev. A}, \bvol{109} (\yr{2024}) \pg{032212}.
		
		\bibitem[52]{catani22}
		{\au{L. Catani}, \au{M. Leifer}, \au{G. Scala}, \au{D. Schmid}, and \au{R. W. Spekkens}}, \at{\textit{What is Nonclassical about Uncertainty Relations?}}, \jt{Phys. Rev. Lett.}, \bvol{129} (\yr{2022}) \pg{240401}.
		
		\bibitem[53]{catani23-2}
		{\au{L. Catani}, \au{M. Leifer}, \au{G. Scala}, \au{D. Schmid}, and \au{R. W. Spekkens}}, \at{\textit{Aspects of the phenomenology of interference that are genuinely nonclassical}}, \jt{Phys. Rev. A}, \bvol{108} (\yr{2023}) \pg{022207}.
		
		\bibitem[54]{barb18}
		{\au{C. Duarte} and \au{B. Amaral}}, \at{\textit{Resource theory of contextuality for arbitrary prepare-and-measure experiments}}, \jt{J. Math. Phys.} (\yr{2018}) \pg{062202}.
		
		\bibitem[55]{regula17}
		{\au{B. Regula}}, \at{\textit{Convex geometry of quantum resource quantification}}, \jt{Journal of Physics A: Mathematical and Theoretical}, \bvol{51}(4) (\yr{2017}) \pg{045303}.
		
		\bibitem[56]{galvao20}
		{\au{E. F. Galvão} and \au{D. J. Brod}}, \at{\textit{``Quantum and classical bounds for two-state overlaps''}}, \jt{Phys. Rev. A}, \bvol{101}(062110) (\yr{2020}).
		
		\bibitem[57]{wagner22A}
		{\au{R. Wagner}, \au{R. S. Barbosa}, and \au{E. F. Galvão}}, \at{\textit{``Inequalities witnessing coherence, nonlocality, and contextuality''}}, \jt{arXiv:2209.02670 [quant-ph]} (\yr{2022}).
		
		\bibitem[58]{chaturvedi21}
		{\au{A. Chaturvedi}, \au{M. Farkas}, and \au{V. J. Wright}}, \at{\textit{Characterising and bounding the set of quantum behaviours in contextuality scenarios}}, \jt{Quantum}, \bvol{5} (\yr{2021}) \pg{484}.
		
		\bibitem[59]{chaturvedi24}
		{\au{G. Viola}, \au{T. Prasad}, \au{E. Panwar}, and \au{A. Chaturvedi}}, , \jt{In preparation.}.
		
		\bibitem[60]{selby20}
		{\au{J. H. Selby} and \au{C. M. Lee}}, \at{\textit{Compositional resource theories of coherence}}, \jt{Quantum}, \bvol{4} (\yr{2020}) \pg{319}.
		
		\bibitem[61]{bruscemi20}
		{\au{D. Schmid}, \au{D. Rosset}, and \au{F. Buscemi}}, \at{\textit{The type-independent resource theory of local operations and shared randomness}}, \jt{Quantum}, \bvol{4} (\yr{2020}) \pg{262}.
		
		\bibitem[62]{zjawin23}
		{\au{B. Zjawin}, \au{D. Schmid}, \au{M. J. Hoban}, and \au{A. B. Sainz}}, \at{\textit{The resource theory of nonclassicality of channel assemblages}}, \jt{Quantum}, \bvol{7} (\yr{2023}) \pg{1134}.
		
		\bibitem[63]{selby23}
		{\au{J. H. Selby}, \au{D. Schmid}, \au{E. Wolfe}, \au{A. B. Sainz}, \au{R. Kunjwal}, and \au{R. W. Spekkens}}, \at{\textit{Accessible fragments of generalized probabilistic theories, cone equivalence, and applications to witnessing nonclassicality}}, \jt{Phys. Rev. A}, \bvol{107} (\yr{2023}) \pg{062203}.
		
		\bibitem[64]{janotta11}
		{\au{P. Janotta}, \au{C. Gogolin}, \au{J. Barrett}, and \au{N. Brunner}}, \at{\textit{Limits on nonlocal correlations from the structure of the local state space}}, \jt{New Journal of Physics}, \bvol{13}(6) (\yr{2011}) \pg{063024}.
		
		\bibitem[65]{muller12}
		{\au{M. P. Müller} and \au{C. Ududec}}, \at{\textit{Structure of Reversible Computation Determines the Self-Duality of Quantum Theory}}, \jt{Phys. Rev. Lett.}, \bvol{108} (\yr{2012}) \pg{130401}.
		
		\bibitem[66]{janotta13}
		{\au{P. Janotta} and \au{R. Lal}}, \at{\textit{Generalized probabilistic theories without the no-restriction hypothesis}}, \jt{Phys. Rev. A}, \bvol{87} (\yr{2013}) \pg{052131}.
		
		\bibitem[67]{barnum19}
		{\au{H. Barnum} and \au{J. Hilgert}}, \at{\textit{Strongly symmetric spectral convex bodies are Jordan algebra state spaces}}, \jt{arXiv.1904.03753 [math-ph]} (\yr{2019}).
	\end{thebibliography}

	\bigskip

	\bigskip
	
	\appendix
	\section{Linear program for testing nonclassicality -- Formal definitions}\label{app1}
	
	The linear program introduced in Ref.~\cite{selby24} is the main tool employed in this manuscript and is constructed entirely in the GPT framework. Here we give some formal definitions for this program and our further modification of what some of the notions mentioned in the main text comprise.
	
	The notion of operational equivalence enables us to define a GPT for the associated operational theory. For this, we concentrate on the quotiented sets $\mathcal{P}/\simeq$ and $K\times\mathcal{M}/\simeq$, i.e., the sets of equivalence classes $[P]$ and $[k|M]$ such that
	\begin{equation}
		[P]:=\{P'\in\mathcal{P}|P'\simeq P\};
	\end{equation}
	\begin{equation}
		[k|M]:=\{k'|M'\in K\times\mathcal{M}|k'|M'\simeq k|M\}.
	\end{equation}
	
	In this framework, every equivalence class $[P]\in\mathcal{P}/\simeq$ in the quotiented operational theory is associated to a state $s_P \in \Omega$, and every equivalence class $[k|M]\in K\times\mathcal{M}/\simeq$ is associated to an effect $e_{k|M} \in \mathcal{E}$. Each state or effect can be described as a vector in a real vector space $V$ equipped with an inner product $\braket{\cdot,\cdot}$. Additionally, these sets must satisfy some specific properties, the most important of which being that the sets $\Omega$ and $\mathcal{E}$ are tomographic. A GPT can be defined as a tuple ($V$,$\braket{\cdot,\cdot}$,$\Omega$,$\mathcal{E}$) satisfying these properties, and a formal definition for the purposes of this paper is given below:
	
	\begin{definition}[GPT]\label{defn:GPT}
		A \emph{generalised probabilistic theory} associated to an operational scenario is a tuple $(\Omega,\mathcal{E},V,\braket{\cdot,\cdot})$ where $\Omega$ and $\mathcal{E}$ are convex sets of vectors in a real vector space $V$ equipped with an inner product $\braket{\cdot,\cdot}$. Moreover,
		\begin{itemize}
			\item \sloppy $\Omega$ does not contain the origin $0\in V$, and each element $s_P\in\Omega$ is associated with an equivalence class ${[P]\in\mathcal{P}/\simeq}$;
			\item $\mathcal{E}$ contains the origin and the privileged \emph{unit} effect $u$, and each element $e_{k|M}\in\mathcal{E}$ is associated with an equivalence class $[k|M]\in K\times\mathcal{M}/\simeq$;
			\item Probabilities are given via the inner product, $p(k|M,P)=\braket{s_P,e_{k|M}}$;
			\item For all $s\in\Omega$, $\frac{1}{\braket{s,u}}s\in\Omega$, where $u$ is the unit effect;
			\item $\Omega$ and $\mathcal{E}$ are \emph{tomographic}, i.e., 
			\begin{equation}
				s_1=s_2\iff\braket{s_1,e}=\braket{s_2,e},\quad\forall e\in\mathcal{E};
			\end{equation}
			\begin{equation}
				e_1=e_2\iff\braket{s,e_1}=\braket{s,e_2},\quad\forall s\in\Omega.
			\end{equation}
		\end{itemize}
	\end{definition}
	
	As mentioned in the main text, however, the linear program in Ref.~\cite{selby24} is written to also take into consideration scenarios in which some of the above properties are not satisfied. In principle, one can simply conceive a relaxation of a GPT in which one merely has access to subsets of the sets $\Omega$ and $\mathcal{E}$. This already implies dropping some features, such as the inclusion of normalised counterparts or tomographic completeness. Such a relaxation receives the name of \emph{GPT fragment}.
	
	\begin{definition}[GPT fragments]\label{defn:fragments}
		A \emph{GPT fragment} associated to an operational scenario is a tuple $(\Omega^F,\mathcal{E}^F,V,\braket{\cdot,\cdot})$ such that $\Omega^F\subseteq\Omega$ and $\mathcal{E}^F\subseteq\mathcal{E}$, with $(\Omega,\mathcal{E},V,\braket{\cdot,\cdot})$ being a GPT.
	\end{definition}
	
	Alternatively, these states and effects can be naturally described with respect to the subspaces of $V$ they span (which will most probably not match\footnote{Consider, for instance, the POM scenario introduced in Sec.~\ref{sec1-3} for $\theta=\frac{\pi}{2}$. In this case, all states lie in the equator of the Bloch sphere, spanning $\mathbb{R}^2$, while the effects remain lying on the three Pauli axes and therefore spanning $\mathbb{R}^3$. This is only one of many examples in which the sets of states and effects in a fragment span different subspaces.}), respectively $\mathsf{Span}(\Omega)$ and $\mathsf{Span}(\mathcal{E})$, and from here on we will always assume this is the case unless stated otherwise. One can easily describe these vectors with respect to the full vector space $V$ by employing inclusion maps $I_\Omega:\mathsf{Span}(\Omega)\to V$ and  $I_\mathcal{E}:\mathsf{Span}(\mathcal{E})\to V$, in which case the tuple $(\Omega,\mathcal{E},I_\Omega,I_\mathcal{E})$ is referred to as the \emph{accessible GPT fragment}~\cite{selby23}. 
	
	\begin{definition}[Accessible GPT fragments]\label{defn:AccessibleFragments}
		An \emph{accessible GPT fragment} associated with an operational scenario is a tuple $(\Omega^A,\mathcal{E}^A,I_\Omega, I_\mathcal{E})$, such that\footnote{In fact, accessible GPT fragments must satisfy more properties than these, for instance, the sets $\Omega^A$ and $\mathcal{E}^A$ must have more structure than merely convexity. We leave these nuances out of this manuscript since they are not relevant to any of the conclusions drawn.}
		\begin{itemize} 
			\item $I_\Omega(\Omega^A)\in V$, $I_\mathcal{E}(\mathcal{E}^A)\in V$;
			\item $p(k|M,P)=\braket{I_\Omega(s),I_\mathcal{E}(e)}$ , for all $[P]\in P/\simeq$ and $[k|M]\in K\times \mathcal{M}/\simeq$;
			\item $(I_\Omega(\Omega^A),I_\mathcal{E},(\mathcal{E}^A),V,\braket{\cdot,\cdot})$ is a GPT fragment.
		\end{itemize}
	\end{definition}
	
	Beyond characterising these sets and the subspaces they span, we can further characterise their positive cones, i.e., the sets $\mathsf{Cone}(\Omega)$ of states that are linear combinations with non-negative coefficients of the states in $\Omega$, and similarly for $\mathsf{Cone}(\mathcal{E})$. In th case that $(\Omega,\mathcal{E})$ are polytopes, we can characterise their cones by the facet inequalities $\{h^\Omega_i\}_{i=1}^n$ for states and $\{h_j^\mathcal{E}\}_{j=1}^m$ for effects, where $h_i^\Omega:\mathsf{span}(\Omega)\to\mathbb{R}$ and $h_j^\mathcal{E}:\mathsf{Span}(\mathcal{E})\to\mathbb{R}$. These are linear maps such that
	\begin{equation}
		h_i^\Omega(v)\geq 0\iff v\in\mathsf{Cone}(\Omega),\,i=1,...,n,
	\end{equation}
	\begin{equation}
		h_j^\mathcal{E}(w)\geq0\iff w\in\mathsf{Cone}(\mathcal{E}),\,j=1,...,m.
	\end{equation}
	Equivalently, we can concatenate the sets $\{h_i^\Omega\}_{i=1}^n$ and $\{h_j^\mathcal{E}\}_{j=1}^m$ into matrices $H_\Omega$, $H_\mathcal{E}$ such that
	\begin{equation}
		H_\Omega(v):=(h_1^{\Omega}(v),...,h_n^{\Omega}(v))^{T}, \forall v\in \mathsf{Span}(\Omega),
	\end{equation}
	\begin{equation}
		H_\mathcal{E}(w):=(h_1^{\mathcal{E}}(w),...,h_n^{\mathcal{E}}(w))^{T}, \forall w \in \mathsf{Span}(\mathcal{E}).
	\end{equation}
	Notice that, by construction,  $H_\Omega(v)$ is entry-wise non-negative iff $v\in\mathsf{Cone}(\Omega)$, and similarly for $H_\mathcal{E}$ iff $w\in\mathsf{Cone}(\mathcal{E})$. Using this characterisation of the accessible GPT fragment, we can introduce the linear program developed in Ref. \cite{selby24}.
	
	\begin{program}\label{program2}   Let $r$ be the minimum depolarising noise that must be added for the statistics obtained by composing any state-effect pair from $\Omega$ and $\mathcal{E}$ to be classically explainable. Given $(H_\Omega,H_\mathcal{E},I_\Omega,I_\mathcal{E})$ characterising the cones of the accessible GPT fragment, $r$ can be computed by the linear program: \\
		\\
		\emph{minimize $r$ such that $$\exists\hspace{0.2cm}\sigma\geq_{e} 0, \text{ an $m\times n$ matrix such that} $$ \begin{equation}\label{code}
				rI_{\mathcal{E}}^{T}\cdot D_{depol}\cdot I_{\Omega}+ (1-r)I_{\mathcal{E}}^{T}\cdot I_{\Omega}=H_{\mathcal{E}}^{T}\cdot \sigma\cdot H_{\Omega}
			\end{equation} where $D_{depol}$ is the completely depolarising channel for the system, and $\geq_e$ denotes entry-wise non-negativity.}
	\end{program}
	
	Notice therefore that the linear program assesses the simplex-embeddability of accessible GPT fragments that are not necessarily within quantum theory. The notion of depolarising noise for such cases is in fact much broader since the code does not demand any particular property from the state labeled as maximally mixed. Abstractly, the code simply ``shrinks'' the set of states towards a point specified by the input maximally mixed state until a simplex embedding becomes possible, which in case of a qubit happens to be the center of the Bloch sphere.
	
	This linear program decides nonclassicality because the existence of a matrix $\sigma$ satisfying equation \ref{code} is equivalent to the existence of a simplex embedding for the GPT fragment depolarised by a factor $r$. This, in turn, has been demonstrated to imply the existence of a noncontextual ontological model for the associated operational scenario~\cite{simplex}. The quantity $r$ therefore may serve as an operational measure of nonclassicality, which we denote as the \emph{robustness of contextuality}. 
	
	\section{Modification of the code}\label{appendix2}
	
	The implementation for the linear program in Ref.~\cite{selby24} requests as inputs the sets of states and effects $(\Omega,\mathcal{E})$ that constitute the fragment, a privileged effect $u$ that plays the role of the \emph{discard} effect and a privileged state $\mu$, called \emph{maximally mixed} state. The code then constructs the accessible GPT fragment associated with the input fragment and the fully depolarising map $D_{depol}$, consisting of discarding any state it receives and preparing the maximally mixed state instead. The code in Mathematica has the additional feature of checking whether the input states and effects are represented by density operators and POVM element matrices, and if they do, already assumes the discard effect to be $\mathbb{1}_\mathcal{H}$ and the maximally mixed state to be $\mathbb{1}_\mathcal{H}/\text{dim}(\mathcal{H})$, where $\mathcal{H}$ is the Hilbert space with respect to which the density operators and POVM elements are represented.\footnote{In this case, the user does not need to provide the discard and the maximally mixed state.} Both implementations then proceed to characterise the accessible GPT fragment and its cone facets and solve Program~\ref{program2}. Finally, they construct an ontological model for the depolarised scenario by factorising the matrix $\sigma$ found by the program, and output an array $(r,\vec{\mu},\vec{\xi})$, where $r$ is the robustness of contextuality to depolarisation, $\vec{\mu}$ is a list of epistemic states for the noncontextual ontological model, and $\vec{\xi}$ a list of the respective response functions.
	
	Ref.~\cite{rossi22} modified the original Mathematica code so that instead of requiring a maximally mixed state as input, it would take a real parameter $\eta\in[0,\pi)$ representing an angle between an axis in the $ZX$ plane of the Bloch sphere and the $Z$ axis. The depolarising map is then replaced by an explicit matrix representation of the fully dephasing map with respect to the axis singled out by $\eta$, and this modification is employed to investigate the interplay between contextuality and coherence in a family of prepare-and-measure scenarios related to the minimum-error state-discrimination task. Although successful for this particular task, this modification is rather limited, since it will not give an account of any GPT fragment whose associated accessible GPT fragment cannot be described in the hemisphere of the Bloch sphere. So even though many important tasks to which contextuality is a resource fit in the scope of this modification, we are interested in a more general approach that can cover the whole scope of quantum prepare-and-measure scenarios.
	
	In the modification introduced by this manuscript, the code asks for the GPT fragment $(\Omega^\mathcal{F},\mathcal{E}^\mathcal{F})$, the discard effect, and a finite collection of effects $\mathcal{M}=\{e_i\}_{i=1}^m$, and then constructs the fully dephasing map as the process that realises the effects $e_i\in\mathcal{E}$ and prepare the respective states $\bar{e}_i\in\Omega$ for which $\braket{\bar{e}_i,e_i}=1$ for all $i=1,...,m$, and where $(V, {\braket{\cdot,\cdot}},\Omega,\mathcal{E})$ is the GPT from which $(\Omega^\mathcal{F},\mathcal{E}^\mathcal{F})$ is derived. Notice therefore that this modification will not give an account of GPTs in which the effects characterising the dephasing axis do not have normalised states as counterparts. In practice, what the code does is to construct the matrix
	\begin{equation}
		D_{deph}=(\bar{e}_1\dots\bar{e}_m)\cdot(e_1\dots e_m)^T,
	\end{equation}
	such that $\bar{e}_i=e_i^T$, and which will replace the depolarising map $D_{depol}$ in Equation~\ref{code}. 
	
	Notice that such implementation is not valid for all possible GPTs, since to conform to Definition~\ref{defn:GPT} does not imply that all effects $\bar{e}_i$ have corresponding states $\bar{e}_i$ as introduced in the previous paragraph. In demanding this structure for the noise model we restrict ourselves to a subset of the \emph{strongly self-dual GPTs}~\cite{janotta11,muller12,janotta13}.
	Given a  GPT $(V,\braket{\cdot,\cdot},\Omega,\mathcal{E})$, and their positive cones
	$\mathsf{Cone}(\Omega)\subset V$ and $\mathsf{Cone}(\mathcal{E})\subset V$. Then:
	\begin{definition}[Strongly self-dual GPT]\label{defn:self-dual} 
		A GPT $(V,\braket{\cdot,\cdot},\Omega,\mathcal{E})$ is \emph{strongly self-dual} if there is a symmetric, positive-semidefinite isomorphism $T:\mathsf{Cone}(\mathcal{E})\to \mathsf{Cone}(\Omega)$ such that, for any $e_1,e_2\in \mathsf{Cone}(\mathcal{E})$, 
		\begin{itemize}
			\item $\braket{T(e_1),e_2}=\braket{T(e_2),e_1}$;
			\item $\braket{T(e_1),e_1}\geq0$.
		\end{itemize}
	\end{definition}
	That is, strongly self-dual GPTs are those in which one can find a symmetric and positive-semidefinite $T$ under which the cones of effects and states are isomorphic $\mathsf{Cone}(\mathcal{E})\simeq \mathsf{Cone}(\Omega)$. Note that GPTs that are not strongly self-dual can be made turned into self-dual GPTs if one restricts the set of effects or states in a suitable way~\cite{janotta13}. In particular, the self-dual theories considered in this work are the ones whose isomorphism $T$ is a transposition of the effect vector. This might miss, for instance, GPTs associated to some Euclidian Jordan algebras~\cite{barnum19}.
	
	Notice also that in principle there is no constraint over the set $\mathcal{M}$ of dephasing effects, for instance, it is not required to form a normalised measurement or to have orthogonal terms. As an additional constraint, the code also checks whether the effects in $\mathcal{M}$ are orthogonal to each other and sum up to the unit effect. This constraint makes sure that, for the quantum case, only the common notion of dephasing is going to be investigated, and these constraints by no means undermine the generality of our results since there is not a clear definition of how a dephasing noise model should look like in GPTs other than quantum theory, so we merely focus our attention to noise models that for sure match the standard notion of quantum dephasing noise. Investigating the noise models ruled out by these constraints, i.e., the ones in which the effects provided by $\mathcal{M}$ are incomplete or coarse-grained, representing projections onto hypervolumes other than an axis, or investigating whether GPTs that are not strongly self-dual accommodate some notion of dephasing noise, are interesting directions for future research. For instance, the Linear Program~\ref{program2} and modifications thereof rely only on properties of the positive cones of states and effects to decide the nonclassicality of GPT fragments. It would be interesting to see if weakly self-dual GPTs\footnote{Weakly self-dual GPTs are the ones whose positive cones of states and effects are isomorphic, with no further restrictions on the isomorphism linking them.} can accommodate some notion of dephasing that still leads to trustworthy statements of simplex-embeddability.
	
	Another aspect of this choice of noise model that deserves attention is that, in principle, the set $\mathcal{M}$ that characterises the dephasing axis does not need to be a subset of $\mathcal{E}$, neither the state counterparts of these effects need to belong in $\Omega$. This might lead to situations in which assessments of nonclassicality in these scenarios become inconclusive since the dephased GPT fragment might not admit of a simplex embedding for any value of $r$. We argue however that allowing for this sort of noise is a reasonable choice that captures the nature of noise in many experiments. Indeed, noise is usually understood as a process replacing the original information encoded in the system with undesired one, which might include information making the experiment more ``nonclassical'' than it is. Furthermore, this is also the approach taken in the formulation of the original Linear Program, in which the maximally mixed state provided is not required to be in $\Omega$~\cite{selby24}. This can also lead to situations in which the whole set of states is ``shrunk'' towards a point that lies outside it, and yet it can capture instances of noise other than depolarisation, such as decay to a (pure) ground state.
	
	Finally, as an example of our reasoning, consider a scenario that takes states and effects from a qubit and in which dephasing happens with respect to the $Z$ axis. The code will receive as input the set $\mathcal{M}=\{\frac{\mathbb{1}+\hat{Z}}{\sqrt2},\frac{\mathbb{1}-\hat{Z}}{\sqrt2}\}$, and the dephasing map takes the form
	\begin{equation}
		D_{deph}=\left(
		\begin{array}{cccc}
			1 & 0 & 0 & 0\\
			0 & 0 & 0 & 0\\
			0 & 0 & 0 & 0\\
			0 & 0 & 0 & 1
		\end{array}\right).
	\end{equation}
	Naturally, this map will act on any state or effect such that
	\begin{equation}
		D_{deph}\cdot\frac{1}{\sqrt2}\left(\begin{array}{c}
			\braket{\mathbb{1}}\\
			\braket{\hat{X}}\\
			\braket{\hat{Y}}\\
			\braket{\hat{Z}}
		\end{array}\right)
		=\frac{1}{\sqrt2}\left(\begin{array}{c}
			\braket{\mathbb{1}}\\
			0\\
			0\\
			\braket{\hat{Z}}
		\end{array}\right),
	\end{equation}
	which is a point in the $Z$ axis of the Bloch sphere.
	
	\section{Numerical investigations}
	
	\subsection{Numerical motivation for the bound \texorpdfstring{$r<\frac12$}{a} for \texorpdfstring{$n$}{b}-to-1 POM with a qubit}\label{appendix3}

	To build up evidence that justifies the bound $r\leq\frac12$ employed in the main text, we once again rely on a numerical investigation using the linear program from Ref.~\cite{selby24}. For this purpose, we consider a prepare-and-measure scenario with an ever-increasing number of equally distributed preparations and measurement outcomes. 
	
	\begin{figure}[h!] 
		\centering
		\includegraphics[scale=0.4]{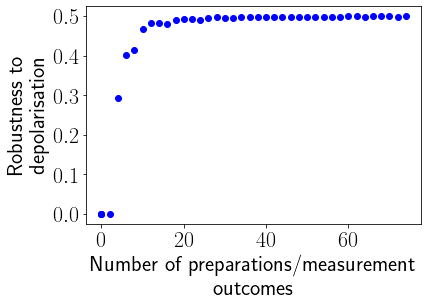}
		\caption{Robustness to depolarisation with the number of preparations and measurement outcomes in the real hemisphere of the Bloch sphere.}
		\label{fig5}
	\end{figure}
	
	For a scenario with $2n$ preparations and measurement outcomes over a qubit Hilbert space, consider the sets
	\begin{equation}
		\Omega_n:=\left\{\ket{k}=\cos\left(\frac{k\pi}{2n}\right)\ket{0}+\sin\left(\frac{k\pi}{2n}\right)\ket{1}\right\}
	\end{equation}
	\begin{equation}
		\mathcal{E}_n:=\left\{\ket{k}=\cos\left(\frac{k\pi}{2n}+\frac{\pi}{8}\right)\ket{0}+\sin\left(\frac{k\pi}{2n}+\frac{\pi}{8}\right)\ket{1}\right\}
	\end{equation}
	with $k=0,...,2n$. For instance, for $n=2$, this example comprises the preparations and measurement outcomes used in the optimal 2-to-1 POM task. Notice that, for $n\to\infty$, this comprises all preparations and measurements in the real hemisphere of the Bloch sphere. We input these sets to the code from Ref.~\cite{selby24} to observe the relation between robustness to depolarisation and the number $2n$ of preparations and outcomes. This plot is given in Figure~\ref{fig5}.
	
	It is immediate to see that robustness quickly approaches the value $\frac12$, from bellow. This alone is enough to build motivation for expecting $r<\frac12$ in a POM task encoding a finite amount of bits into a finite amount of preparations since these values will only be achievable for $n\gg3$ preparations and measurements.
	
	As argued in the main text, one can easily generalise this method of inspection to quantum systems of greater dimension, recovering bounds of robustness of contextuality and consequently, of optimal quantum performance for parity-oblivious multiplexing tasks via Eq.~\ref{robustness1}.
	
	\subsection{Numerical motivation for neglecting dephasing axes beyond $X$, $Y$ and $Z$}\label{app4}
	
	In Sec.~\ref{sec3}, we chose minimising robustness to dephasing over the axes $X$, $Y$, and $Z$. We here show by a numerical investigation that indeed the axes $X$, $Y$, and $Z$ are the only ones that matter.
	
	By using a pseudo-random number generator function, we can generate random axes of the form
	\begin{equation}
		\left\{\frac{1}{\sqrt2}\left(\begin{array}{c}
			1\\
			\sin\theta_r\cos\phi_r\\
			\sin\theta_r\sin\phi_r\\
			\cos\theta_r
		\end{array}\right),\frac{1}{\sqrt2}\left(\begin{array}{c}
			1\\
			-\sin\theta_r\cos\phi_r\\
			-\sin\theta_r\sin\phi_r\\
			-\cos\theta_r
		\end{array}\right)\right\},
	\end{equation}
	where $\theta_r$ and $\phi_r$ are independently generated random angles.
	
\begin{figure}[h!]
		\centering
		\includegraphics[scale=0.4]{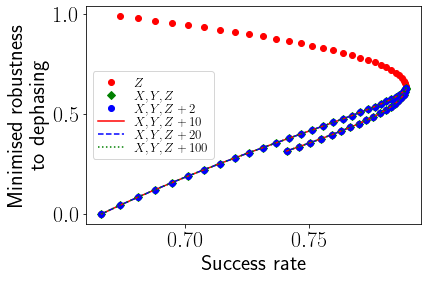}
		\caption{Robustness to dephasing vs. success rate for the 3-to-1 POM task, minimised over axes $X$, $Y$, $Z$ and additional pseudo-random axes.}
		\label{fig6}
	\end{figure}

	We then compute the relation between robustness to dephasing and success rate with respect to the $Z$ axis; minimised over $X$, $Y$ and $Z$ axis; minimised over $X$, $Y$, $Z$ and $n$ random axes, with $n$ being 2, 10, 20 and 100, respectively. The plot is displayed in Figure~\ref{fig6}.
	
	Notice that once we minimise robustness to dephasing over axes $X$, $Y$, and $Z$, the addition of the extra random axes becomes irrelevant. Indeed, all points for the plots with extra axes coincide with the plots without it. Despite this not being a minimisation over all possible axes, it provides enough evidence for our conjecture to hold.
	
\end{document}